%% file: DIS2016-JuanRojo-NNPDF.tex
\def\gsim{\mathrel{\rlap{\lower4pt\hbox{\hskip1pt$\sim$}}
    \raise1pt\hbox{$>$}}}         %greater than or approx. symbol
\def\lsim{\mathrel{\rlap{\lower4pt\hbox{\hskip1pt$\sim$}}
    \raise1pt\hbox{$<$}}}         %less than or approx. symbol

\documentclass{PoS}

\title{Weak boson production from D0 and LHCb
in the NNPDF global analysis}

\ShortTitle{Progress in the NNPDF global analysis}

\author{\speaker{Juan Rojo}~\footnote{Presented on behalf of the NNPDF Collaboration.}\\
       University of Oxford\\
       E-mail: \email{juan.rojo@physics.ox.ac.uk}}

\abstract{High-precision electroweak production measurements from the
  Tevatron and the LHC provide important constraints on the quark
  flavor separation in global PDF fits.
  In this contribution, we study the impact of the recent D0 
  $W$ asymmetry measurements and of the LHCb $W$ and $Z$  Run I combination
  in the global NNPDF analysis.
  We find that these measurements can be described by NLO QCD theory
  and that they lead to a significant reduction of PDF uncertainties.
}

\FullConference{XXIV International Workshop on Deep-Inelastic Scattering and Related Subjects\\
         11-15 April, 2016\\
         DESY Hamburg, Germany}

\begin{document}

\paragraph{Towards NNPDF3.1.}

NNPDF3.0~\cite{Ball:2014uwa} is the most
updated version of the NNPDF family of global analysis.
NNPDF3.0 was the result of a complete rewrite of the NNPDF fitting code from
{\tt Fortran} to {\tt C++} and {\tt Python}, and for the first time
the entire fitting methodology was
robustly validated by means of closure tests.
Subsequently, a number of studies based on NNPDF3.0 have been presented,
including the constraints on
NNPDF3.0 from the final HERA I+II combined inclusive dataset~\cite{Rojo:2015nxa}
and those from  forward charm production at LHCb~\cite{Gauld:2015yia}.
The NNPDF3.0 fits have also been the baseline to construct PDF sets
with NLO+NLL and NNLO+NNLL threshold resummation~\cite{Bonvini:2015ira},
used as input for updated NLO+NLL calculations
of squark and gluino production at the LHC 13 TeV~\cite{Beenakker:2015rna}.
NNPDF3.0 is also part of the PDF4LHC recommendations for PDF usage
at Run II~\cite{Butterworth:2015oua}.

In this contribution, we review recent progress towards the next major
update of the
NNPDF family of global analysis, NNPDF3.1.
First of all, we review the new fit settings, including
improvements in data, theory and methodology,
and then study the impact of some selected
datasets.
In particular, we will
consider the impact of the D0 legacy measurements on $W$ leptonic asymmetries
and of the LHCb combination of $W$ and $Z$ production in the muon
final state from Run I.
We will show that these datasets provide important constraints in the global
analysis, in particular regarding quark flavor separation.

\paragraph{Data, theory and fitting methodology.}

As compared to the NNPDF3.0 analysis, the NNPDF3.1 fit introduces a number of
improvements in terms of the input
dataset, theory calculations
and the fitting methodology.
Concerning the fitted dataset, in addition to the combined HERA inclusive
data, NNPDF3.1 will include, among others, the Tevatron D0 Run II $W$ lepton
asymmetries, the complete Run I $W,Z$ production dataset
from LHCb, new inclusive jet and electroweak
production measurements from ATLAS and CMS, as well as the differential
distributions for top pair quark production  and the $Z$ transverse
momentum at 8 TeV from ATLAS and CMS, exploiting recent progress in
NNLO calculations~\cite{Czakon:2015owf,Ridder:2016nkl,Boughezal:2015ded}.

From the point of view of the theory calculations,
structure functions and PDF evolution are now evaluated
using the
public PDF evolution code {\tt APFEL}~\cite{Bertone:2013vaa}, suitably
benchmarked with the internal code {\tt FKgenerator}~\cite{Ball:2010de}
used
in previous NNPDF fits.
Among other recent improvements, {\tt APFEL} allows for calculations
using both the pole and the running heavy quark mass schemes, as well as
the calculation of
FONLL general-mass scheme structure functions with massive charm-initiated
contributions~\cite{Ball:2015tna}.

In terms of the fitting methodology, the main difference as compared
to NNPDF3.0 is the treatment of the charm PDF on an equal
footing as the light quark PDFs and the gluon, following
the strategy presented in Ref~\cite{Ball:2016neh}.
Fitting the charm PDF has a number of conceptual advantages, including 
an increased stability of the PDFs
with respect the value of the charm mass
$m_c$.
Other updates include an improved treatment of the PDF positivity
constraints, specially relevant for the production of BSM
high-mass resonances, as well as a more refined determination
of the asymptotic behaviour of PDFs at small and large-$x$
relevant for the determination of the preprocessing ranges~\cite{Ball:2016spl}.

\paragraph{The impact of the D0 Run II $W$ lepton asymmetry data.}
The D0 collaboration at the Tevatron has presented their legacy measurements for the $W$ leptonic
asymmetries in the electron~\cite{D0:2014kma} and muon~\cite{Abazov:2013rja} final states
using the complete Run II luminosity.
The impact of these data was demonstrated in the HERAfitter (now xFitter)
analysis of Ref.~\cite{Camarda:2015zba}, and
some of these D0 measurements are already included in other
 global PDF fits.
 Here we show the impact of the D0 $W$ asymmetry data when added on top
 of a baseline fit using the NNPDF3.1 settings, but with the NNPDF3.0
 dataset.
 NLO QCD cross-sections have been computed with {\tt MCFM} interfaced
 to {\tt APPLgrid}~\cite{Carli:2010rw},
 using the same settings as in~\cite{Camarda:2015zba}.

%%%%%%%%%%%%%%%%%%%%%%%%%%%%%%%%%%%%%%%%%%%%%%%%%%%%%%%%%%%%%%%%%%%%%%%%%%%%%%%%%%%%%%%%%%
\begin{figure}[t]
\centering
\includegraphics[width=0.49\textwidth]{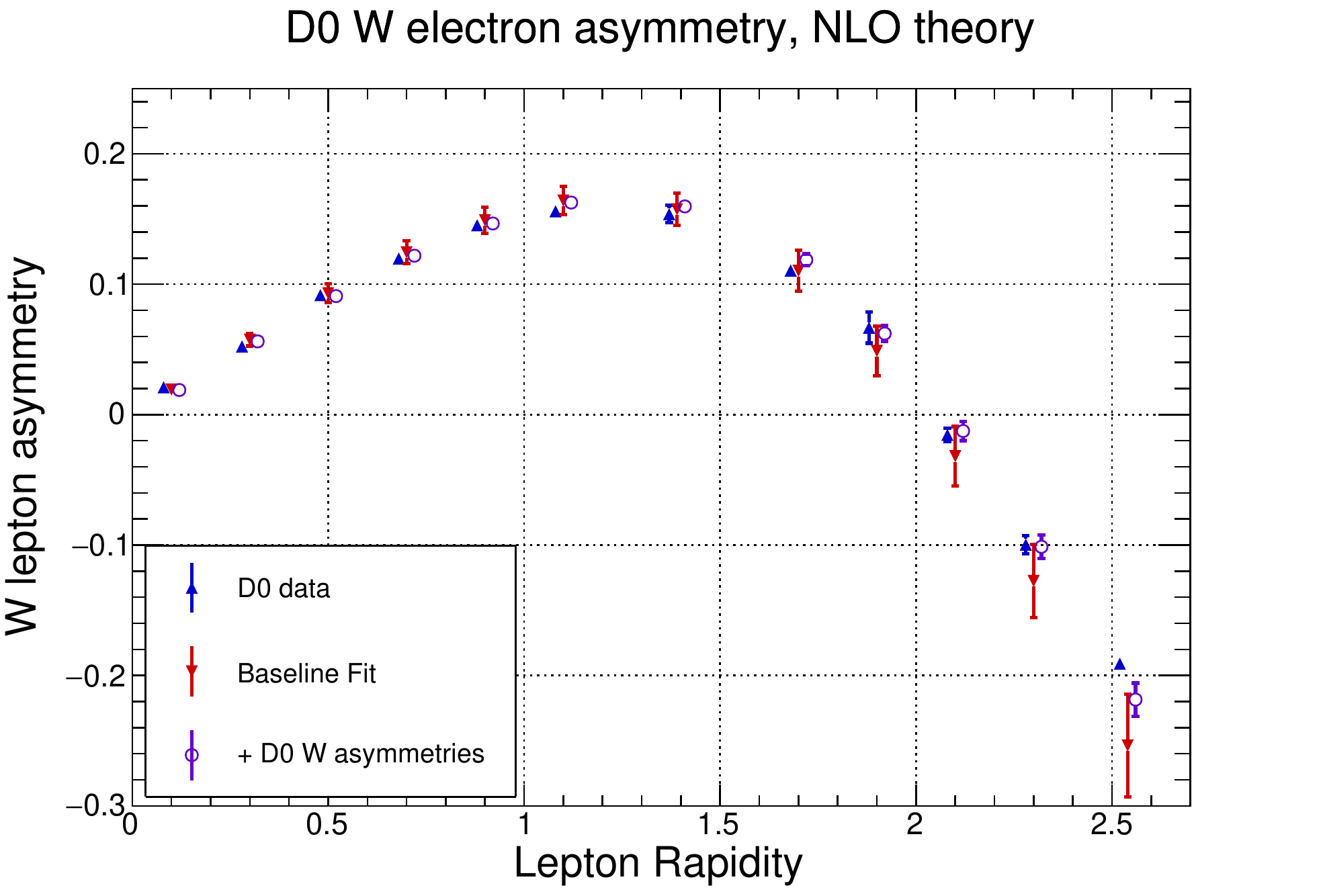}
\includegraphics[width=0.49\textwidth]{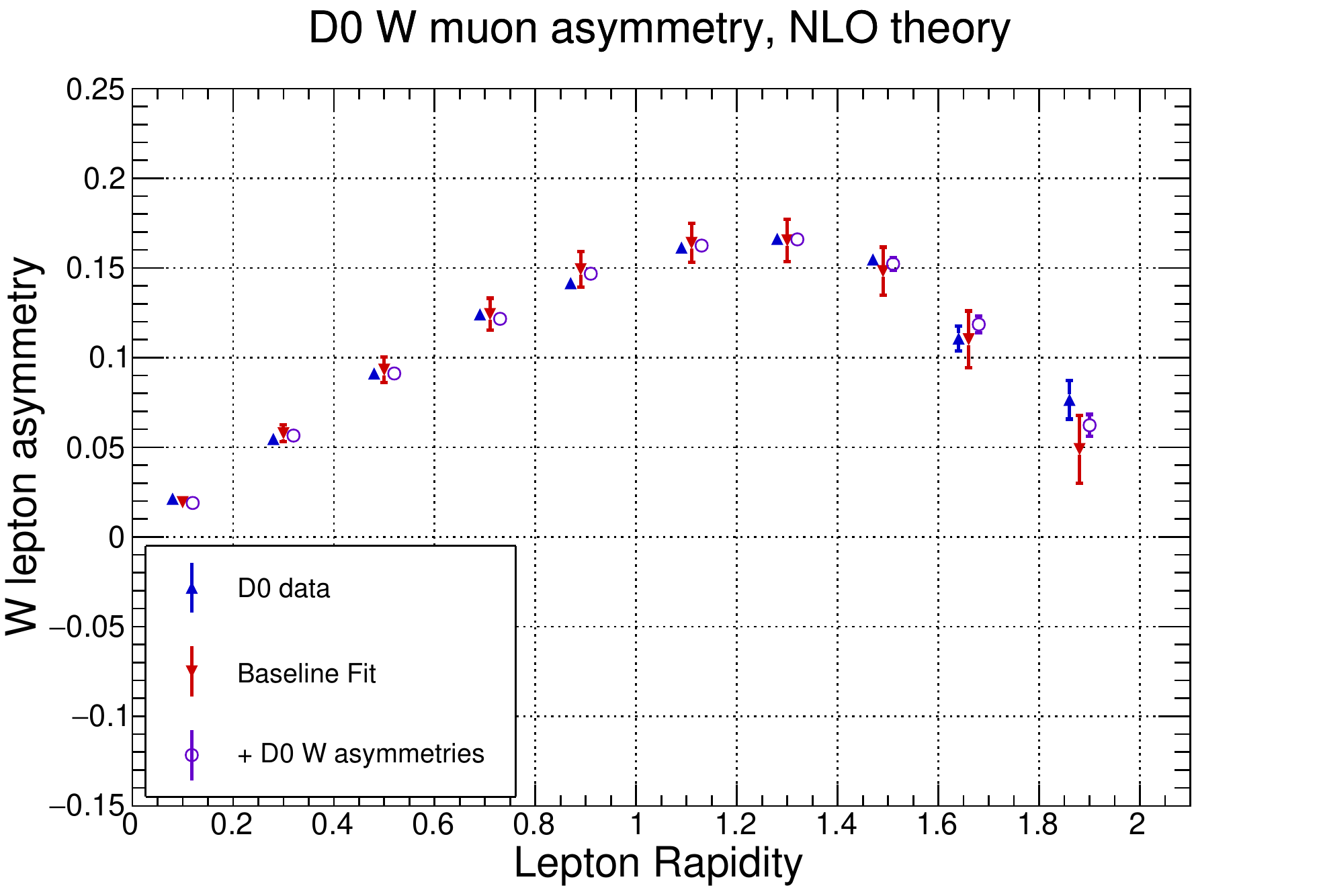}
\includegraphics[width=0.49\textwidth]{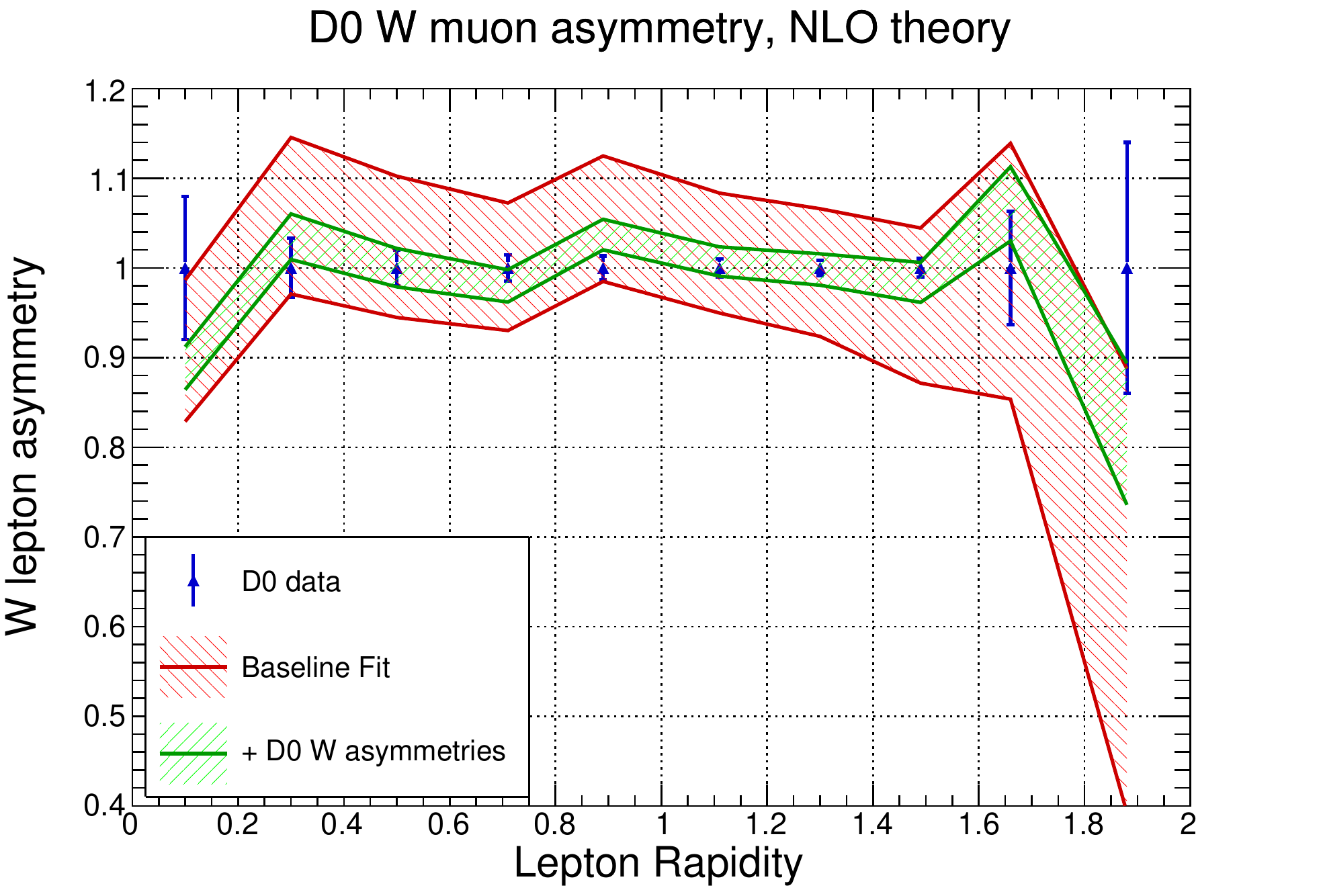}
\includegraphics[width=0.49\textwidth]{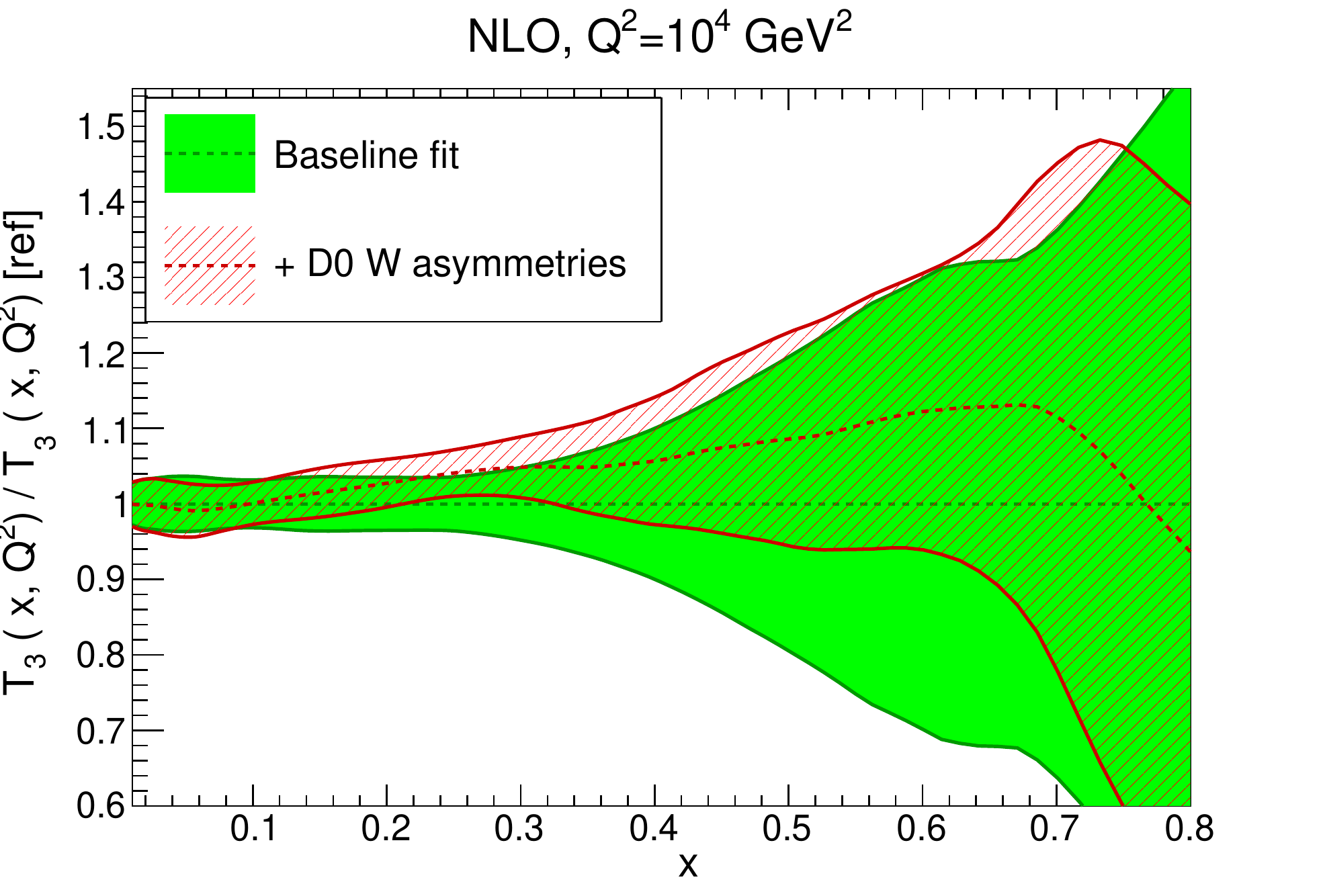}
\vspace{-0.2cm}
\caption{\small Upper plots: comparison between data and NLO QCD theory for the D0 electron (left)
  and muon (right) $W$ asymmetries as a function of the lepton rapidity.
  The experimental data is compared with a baseline fit, with NNPDF3.1 methodology and NNPDF3.0
  dataset, as well as with the
  same fit now including the D0 $W$ asymmetries.
  Lower left plot: the ratio between theory and data for the D0 $W$ muon asymmetry
  data.
  Lower right plot: the isotriplet quark PDF, $T_3=u+\bar{u}-d-\bar{d}$, in the baseline fit and in the fit with
  the D0 data included, normalized to the central value of the former.}
\label{fig:TevatronLegacyData}
\end{figure}
%%%%%%%%%%%%%%%%%%%%%%%%%%%%%%%%%%%%%%%%%%%%%%%%%%%%%%%%%%%%%%%%%%%%%%%%%%%%%%%%%%%%%%%%%%

In Fig.~\ref{fig:TevatronLegacyData} we show a
comparison between the D0 data and NLO theory for the electron 
  and muon $W$ asymmetries, as a function of the lepton rapidity.
  The experimental data is compared with the baseline fit
  and with the same fit now including the D0 $W$ asymmetries.
  The PDF uncertainties in the theory predictions are substantially reduced once the D0
  data is included in the fit, highlighting the constraining power of these measurements.
  In Fig.~\ref{fig:TevatronLegacyData} we also show
 the same comparisons for the muon asymmetry, this time normalizing
 the theory predictions to the experimental data.
  This illustrates the substantial reduction of PDF uncertainties
  in the entire kinematical range, but specially for forward
  rapidities, sensitive to the poorly-known large-$x$
  antiquarks.
  The good agreement between data and theory after the fit is
  also indicated by the $\chi^2/N_{\rm dat}$ estimator, which
  is $\simeq 1.5$ for the electron data and $\simeq 1.4$
  for the muon data.

  In Fig.~\ref{fig:TevatronLegacyData} we also show the impact of the D0 $W$ asymmetries
  on the isotriplet quark PDF combination, $T_3=u+\bar{u}-d-\bar{d}$,
  for $Q^2=100$ GeV$^2$.
  The most significant effect appears to be for $x\gsim 0.3$, where the
  D0 data prefer a harder central value for $T_3$, as well as a 
  reduction of the PDF uncertainties that can be as large as 50\%,
  for example for $x\simeq 0.6$.
  Similar constraints are observed for other quark combinations.

\paragraph{The LHCb Run I $W,Z$ production combination.}

%%%%%%%%%%%%%%%%%%%%%%%%%%%%%%%%%%%%%%%%%%%%%%%%%
\begin{figure}[h!]
\centering
\includegraphics[width=0.49\textwidth]{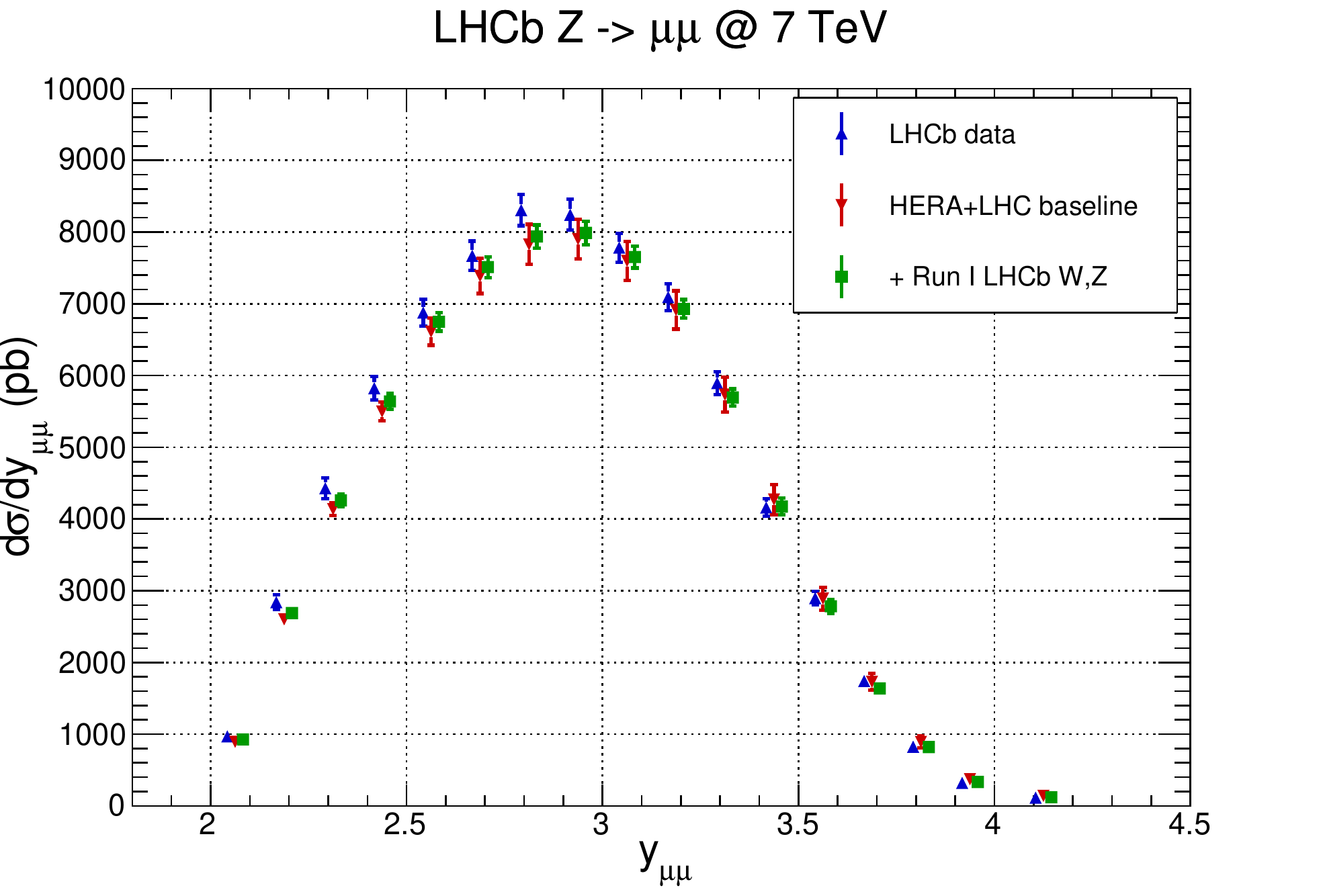}
\includegraphics[width=0.49\textwidth]{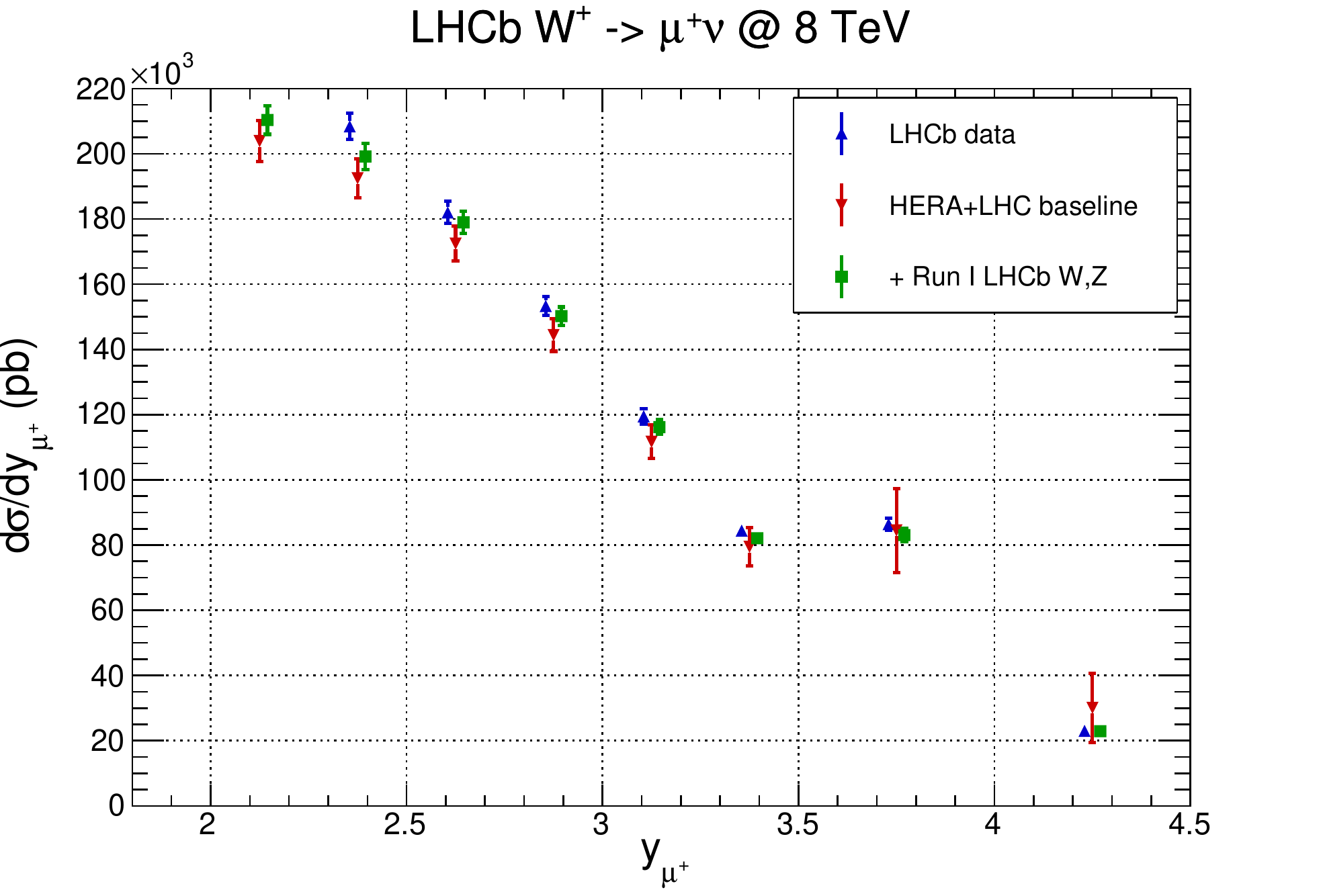}
\includegraphics[width=0.49\textwidth]{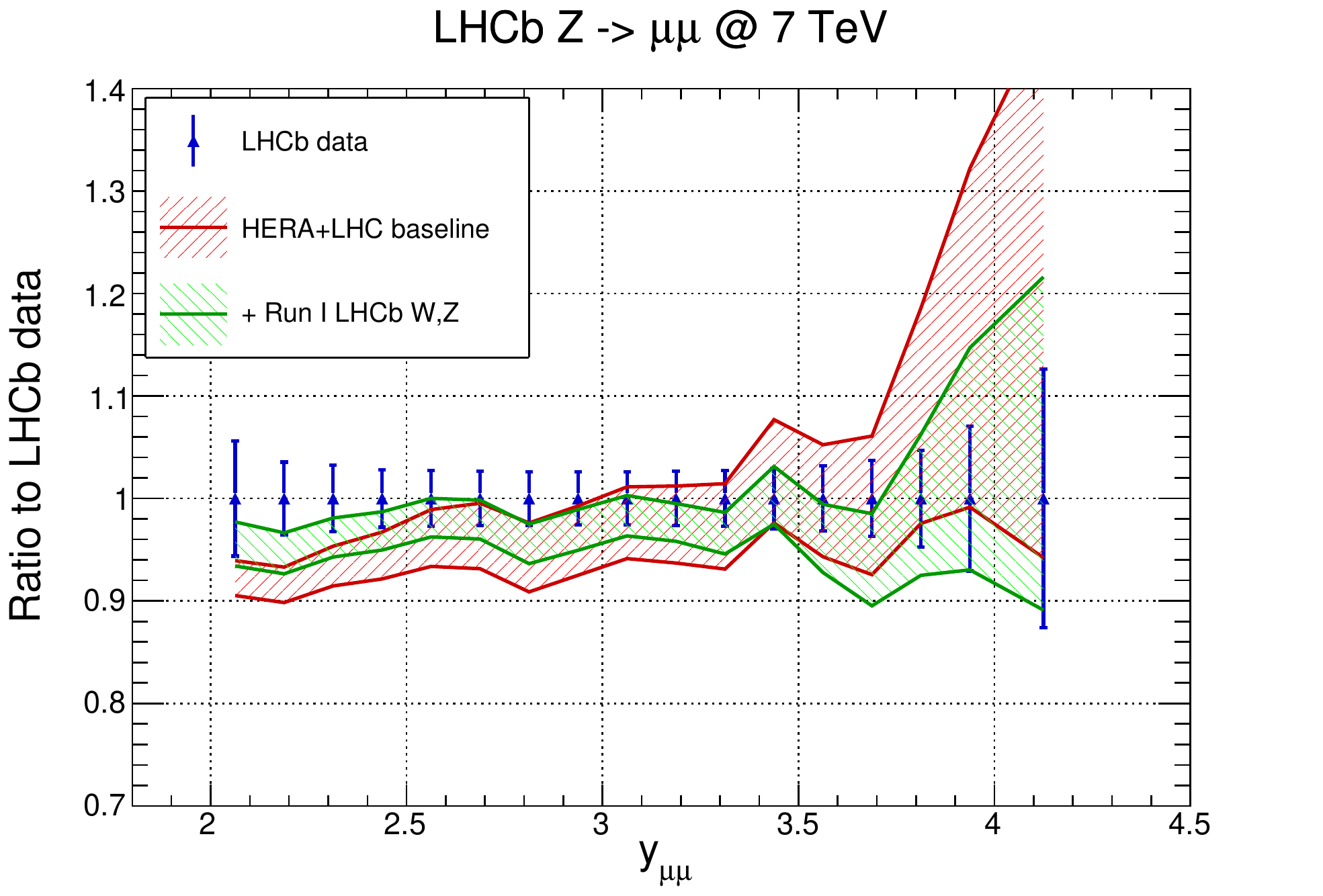}
\includegraphics[width=0.49\textwidth]{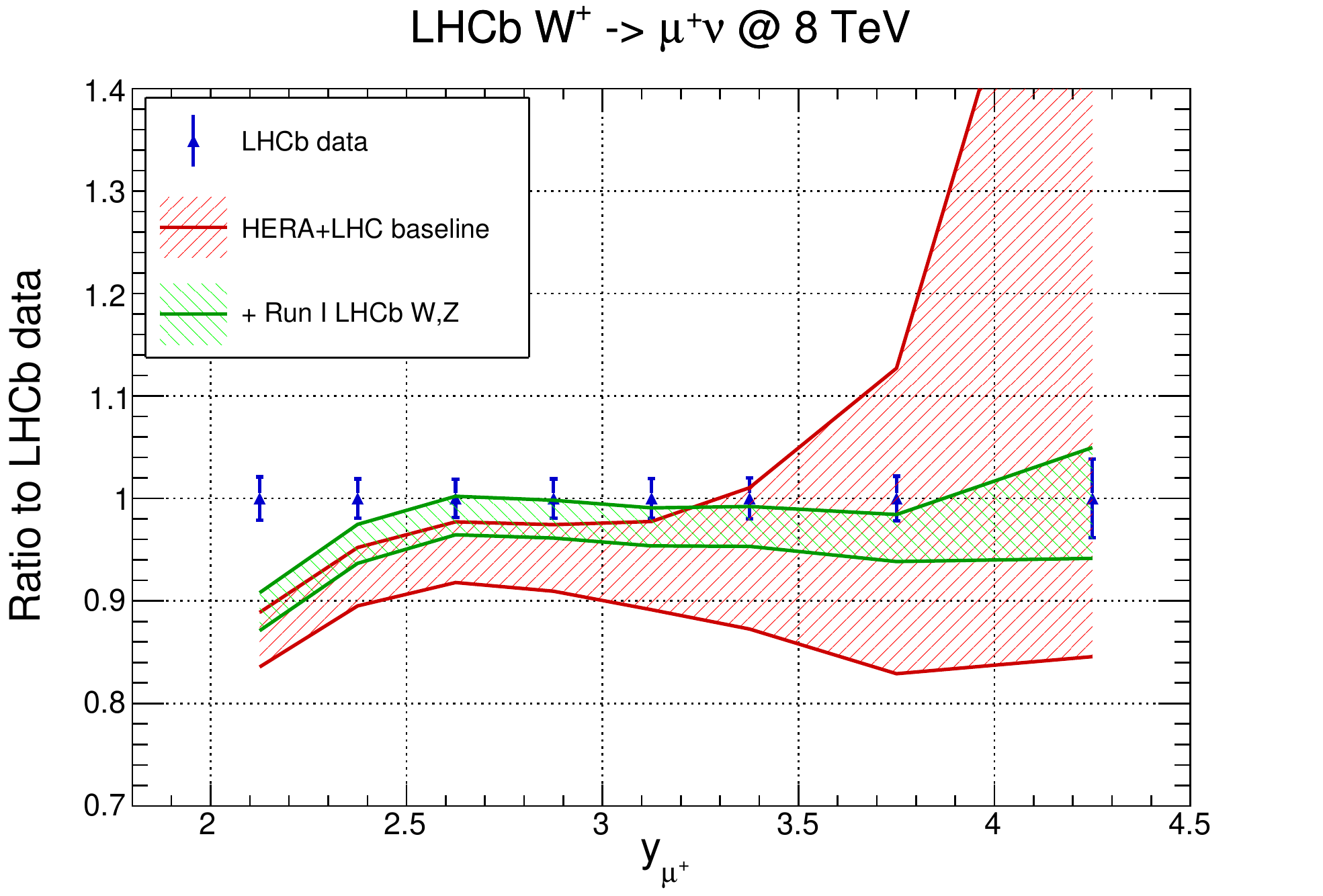}
\vspace{-0.2cm}
\caption{\small Upper plots: the differential cross-sections
  for $Z$ production at 7 TeV and $W^+$ production at 8 TeV
  in the muon final state, measured
 by the LHCb experiment in the forward
  region.
We compare a baseline HERA+LHC NLO QCD fit to the same
  fit once the LHCb data has been included.
  Lower plots: same comparison, now normalized to the
  LHCb measurements.
}
\label{fig:LHCbLegacyData}
\end{figure}
%%%%%%%%%%%%%%%%%%%%%%%%%%%%%%%%%%%%%%%%%%%%%%%%%

Now we discuss the impact of the recent LHCb electroweak gauge
boson production data in the forward region at 7 TeV~\cite{Aaij:2015gna} and
8 TeV~\cite{Aaij:2015zlq} in the muon final state.
These measurements are provided including the full experimental covariance
matrix with the correlations between the $Z^0$, $W^+$ and $W^-$ rapidity
distributions at the two center of mass energies.
They supersede previous LHCb Run I measurements of $W,Z$ production
in the muon final state, and  represent
the LHCb legacy measurements at 7 and 8 TeV.
For the study of the impact of these datasets,
the baseline  PDF fit is provided by a HERA+LHC fit,
including the same LHC experiments as in NNPDF3.0.
Then on top of this we have added the LHCb 7 and 8 TeV combined
$W$ and $Z$ data.
As in the case of the D0 $W$ asymmetries,
NLO QCD predictions have been computed using {\tt MCFM} interfaced
to {\tt APPLgrid}.

In Fig.~\ref{fig:LHCbLegacyData} we show, in the
upper plots, the differential cross-sections
  for $Z$ production at 7 TeV and $W^+$ production at 8 TeV.
  We compare the baseline HERA+LHC fit to the same
  fit once the LHCb data has been included.
  The lower plots then
  represent the same comparison, now  normalized to the
  LHCb measurements.
  As we can see, the agreement between data and theory is
  in general good after the fit,
  and a substantial reduction of the PDF uncertainties is obtained, specially
  for the charged-current Drell-Yan data.
  The only exception is the most central rapidity bin at 8 TeV,
  which seems to overshoot
  the theory for all final states ($Z$, $W^+$ and $W^-$).

\begin{figure}[t]
\centering
\includegraphics[width=0.49\textwidth]{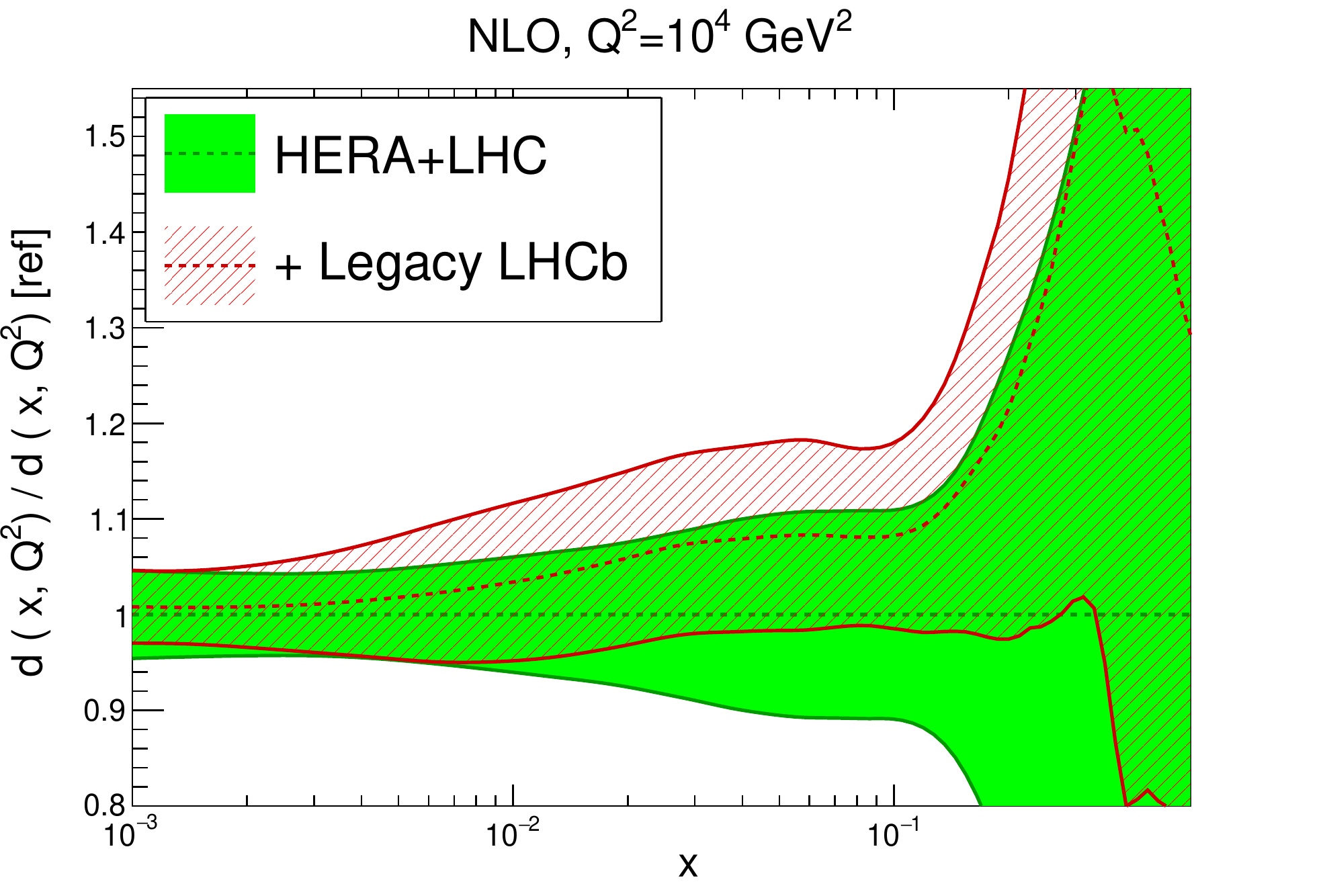}
\includegraphics[width=0.49\textwidth]{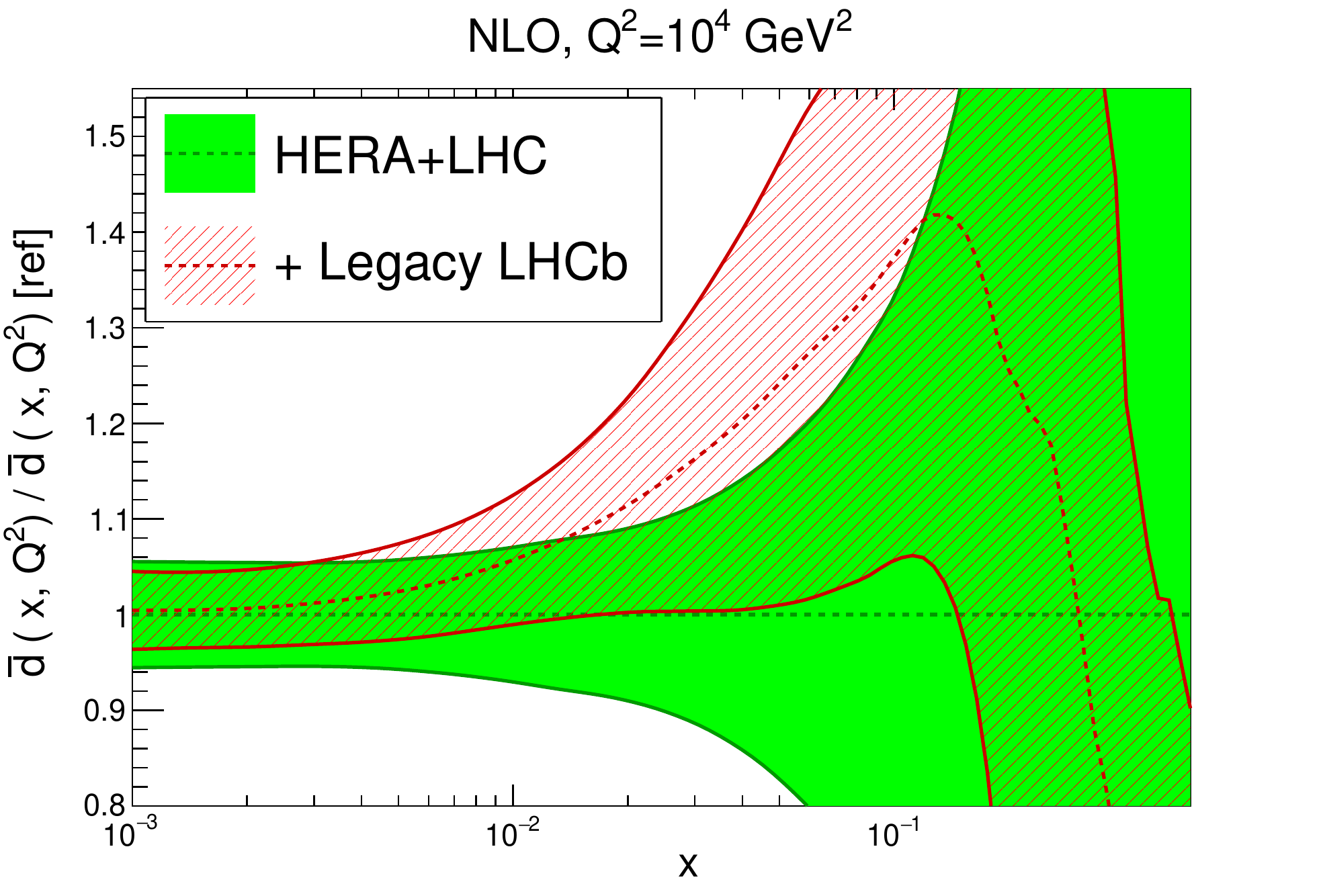}
\vspace{-0.2cm}
\caption{\small Comparison of the down (left) and anti-down (right plot) quark
PDFs between the HERA+LHC baseline fit and the same fit including the LHCb Run I
$W$ and $Z$ production data.
Results are shown at $Q^2=100$ GeV$^2$, normalized to the central value of
the baseline fit.
}
\label{fig:LHCbLegacyDataPDFs}
\end{figure}

Then in Fig.~\ref{fig:LHCbLegacyDataPDFs} we show
a comparison of the down (left) and anti-down (right plot) quark
PDFs between the HERA+LHC baseline fit and the same fit including the LHCb Run I
$W$ and $Z$ production data.
Results are shown at $Q^2=100$ GeV$^2$, normalized to the central value of
the baseline fit.
We find that the LHCb Run I electroweak data prefer a harder down and anti-down
quarks at medium and large-$x$,  except for $x\gsim 0.4$ where the anti-down
quark tends to become smaller than in the baseline.

%%%%%%%%%%%%%%%%%%%%%%%%%%%%%%%%%%%%%%%%%%%%%%%%%
%%%%%%%%%%%%%%%%%%%%%%%%%%%%%%%%%%%%%%%%%%%%%%%%%

\paragraph{Outlook.}
NNPDF3.1 is the forthcoming update of the NNPDF family of global
analysis.
Significant improvements in terms of experimental data, theory
calculations, and fitting methodology have been implemented.
Here we have discussed the significant constraints that the updated
D0 and LHCb electroweak production measurements appear to
provide in the global PDF fit.
These measurements help disentangling
the quarks and anti-quarks of different
flavor, and thus represent a useful addition to the next
generation of global PDF analysis.

\paragraph{Acknowledgments.}
This work has been supported by an STFC Rutherford Fellowship
and Grant ST/K005227/1 and ST/M003787/1, and
by an European Research Council Starting Grant ``PDF4BSM''.
We thank S.~Camarda for providing us with the {\tt applgrids}
of Ref.~\cite{Camarda:2015zba}.

\input{DIS2016-JuanRojo-NNPDF.bbl}
%\bibliography{DIS2016-JuanRojo-NNPDF}

\end{document}

%% file: DIS2016-JuanRojo-NNPDF.bbl
\providecommand{\href}[2]{#2}\begingroup\raggedright\endgroup

%% file: DIS2016-JuanRojo-NNPDF.bbl
\begin{thebibliography}{10}

\bibitem{Ball:2014uwa}
{\bf NNPDF} Collaboration, R.~D. Ball et~al., {\it {Parton distributions for
  the LHC Run II}},  {\em JHEP} {\bf 04} (2015) 040,
  [\href{http://arxiv.org/abs/1410.8849}{{\tt arXiv:1410.8849}}].

\bibitem{Rojo:2015nxa}
J.~Rojo, {\it {Progress in the NNPDF global analysis and the impact of the
  legacy HERA combination}},  {\em PoS} {\bf EPS-HEP2015} (2015) 506,
  [\href{http://arxiv.org/abs/1508.07731}{{\tt arXiv:1508.07731}}].

\bibitem{Gauld:2015yia}
R.~Gauld, J.~Rojo, L.~Rottoli, and J.~Talbert, {\it {Charm production in the
  forward region: constraints on the small-x gluon and backgrounds for neutrino
  astronomy}},  {\em JHEP} {\bf 11} (2015) 009,
  [\href{http://arxiv.org/abs/1506.08025}{{\tt arXiv:1506.08025}}].

\bibitem{Bonvini:2015ira}
M.~Bonvini, S.~Marzani, J.~Rojo, L.~Rottoli, M.~Ubiali, R.~D. Ball, V.~Bertone,
  S.~Carrazza, and N.~P. Hartland, {\it {Parton distributions with threshold
  resummation}},  {\em JHEP} {\bf 09} (2015) 191,
  [\href{http://arxiv.org/abs/1507.01006}{{\tt arXiv:1507.01006}}].

\bibitem{Beenakker:2015rna}
W.~Beenakker, C.~Borschensky, M.~Kramer, A.~Kulesza, E.~Laenen, S.~Marzani, and
  J.~Rojo, {\it {NLO+NLL squark and gluino production cross-sections with
  threshold-improved parton distributions}},  {\em Eur. Phys. J.} {\bf C76}
  (2016), no.~2 53, [\href{http://arxiv.org/abs/1510.00375}{{\tt
  arXiv:1510.00375}}].

\bibitem{Butterworth:2015oua}
J.~Butterworth et~al., {\it {PDF4LHC recommendations for LHC Run II}},  {\em J.
  Phys.} {\bf G43} (2016) 023001, [\href{http://arxiv.org/abs/1510.03865}{{\tt
  arXiv:1510.03865}}].

\bibitem{Czakon:2015owf}
M.~Czakon, D.~Heymes, and A.~Mitov, {\it {High-precision differential
  predictions for top-quark pairs at the LHC}},  {\em Phys. Rev. Lett.} {\bf
  116} (2016), no.~8 082003, [\href{http://arxiv.org/abs/1511.00549}{{\tt
  arXiv:1511.00549}}].

\bibitem{Ridder:2016nkl}
A.~Gehrmann-De~Ridder, T.~Gehrmann, E.~W.~N. Glover, A.~Huss, and T.~A. Morgan,
  {\it {The NNLO QCD corrections to Z boson production at large transverse
  momentum}},  \href{http://arxiv.org/abs/1605.04295}{{\tt arXiv:1605.04295}}.

\bibitem{Boughezal:2015ded}
R.~Boughezal, J.~M. Campbell, R.~K. Ellis, C.~Focke, W.~T. Giele, X.~Liu, and
  F.~Petriello, {\it {Z-boson production in association with a jet at
  next-to-next-to-leading order in perturbative QCD}},  {\em Phys. Rev. Lett.}
  {\bf 116} (2016), no.~15 152001, [\href{http://arxiv.org/abs/1512.01291}{{\tt
  arXiv:1512.01291}}].

\bibitem{Bertone:2013vaa}
V.~Bertone, S.~Carrazza, and J.~Rojo, {\it {APFEL: A PDF Evolution Library with
  QED corrections}},  {\em Comput.Phys.Commun.} {\bf 185} (2014) 1647,
  [\href{http://arxiv.org/abs/1310.1394}{{\tt arXiv:1310.1394}}].

\bibitem{Ball:2010de}
{\bf {The NNPDF }} Collaboration, R.~D. Ball et~al., {\it {A first unbiased
  global NLO determination of parton distributions and their uncertainties}},
  {\em Nucl. Phys.} {\bf B838} (2010) 136,
  [\href{http://arxiv.org/abs/1002.4407}{{\tt arXiv:1002.4407}}].

\bibitem{Ball:2015tna}
R.~D. Ball, V.~Bertone, M.~Bonvini, S.~Forte, P.~Groth~Merrild, J.~Rojo, and
  L.~Rottoli, {\it {Intrinsic charm in a matched general-mass scheme}},  {\em
  Phys. Lett.} {\bf B754} (2016) 49--58,
  [\href{http://arxiv.org/abs/1510.00009}{{\tt arXiv:1510.00009}}].

\bibitem{Ball:2016neh}
{\bf The NNPDF} Collaboration, R.~D. Ball, V.~Bertone, M.~Bonvini, S.~Carrazza,
  S.~Forte, A.~Guffanti, N.~P. Hartland, J.~Rojo, and L.~Rottoli, {\it {A
  Determination of the Charm Content of the Proton}},
  \href{http://arxiv.org/abs/1605.06515}{{\tt arXiv:1605.06515}}.

\bibitem{Ball:2016spl}
R.~D. Ball, E.~R. Nocera, and J.~Rojo, {\it {The asymptotic behaviour of parton
  distributions at small and large $x$}},
  \href{http://arxiv.org/abs/1604.00024}{{\tt arXiv:1604.00024}}.

\bibitem{D0:2014kma}
V.~M. Abazov et~al., {\it Measurement of the electron charge asymmetry in
                        $p\bar{p}\rightarrow W+X \rightarrow e\nu +X$
                        decays in $p\bar{p}$ collisions at
                        $\sqrt{s}=1.96$ TeV},  {\em Phys.Rev.}
{\bf D91} (2015) 032007,
  [\href{http://arxiv.org/abs/1412.2862}{{\tt arXiv:1412.2862}}].

\bibitem{Abazov:2013rja}
{\bf D0} Collaboration, V.~M. Abazov et~al., {\it {Measurement of the muon
  charge asymmetry in $p\bar{p}$ $\to$ W+X $\to$ $\mu$$\nu$ + X events at
  $\sqrt{s}$=1.96 TeV}},  {\em Phys.Rev.} {\bf D88} (2013) 091102,
  [\href{http://arxiv.org/abs/1309.2591}{{\tt arXiv:1309.2591}}].

\bibitem{Camarda:2015zba}
{\bf HERAFitter developers' Team}, S.~Camarda et~al., {\it {QCD
    analysis of $W$- and $Z$-boson production at Tevatron}},
{\em Eur.\ Phys.\ J.\ C} {\bf 75}, no. 9, 458 (2015),
  [\href{http://arxiv.org/abs/1503.05221}{{\tt arXiv:1503.05221}}].

\bibitem{Carli:2010rw}
T.~Carli et~al., {\it {A posteriori inclusion of parton density functions in
  NLO QCD final-state calculations at hadron colliders: The APPLGRID Project}},
   {\em Eur.Phys.J.} {\bf C66} (2010) 503,
  [\href{http://arxiv.org/abs/0911.2985}{{\tt arXiv:0911.2985}}].

\bibitem{Aaij:2015gna}
{\bf LHCb} Collaboration, R.~Aaij et~al., {\it {Measurement of the forward $Z$
  boson production cross-section in $pp$ collisions at $\sqrt{s}=7$ TeV}},
  {\em JHEP} {\bf 08} (2015) 039, [\href{http://arxiv.org/abs/1505.07024}{{\tt
  arXiv:1505.07024}}].

\bibitem{Aaij:2015zlq}
{\bf LHCb} Collaboration, R.~Aaij et~al., {\it {Measurement of forward W and Z
  boson production in $pp$ collisions at $ \sqrt{s}=8 $ TeV}},  {\em JHEP} {\bf
  01} (2016) 155, [\href{http://arxiv.org/abs/1511.08039}{{\tt
  arXiv:1511.08039}}].

\end{thebibliography}
